\documentclass{ws-ijmpd}

\begin{document}

\markboth{I. Georgantopoulos}
{Compton-thick AGN}

%
\catchline{}{}{}{}{}
%

\title{Recent developments  in the search for Compton-thick AGN}

\author{I. Georgantopoulos}

\address{Osservatorio Astronomico di Bologna, INAF, Via Ranzani 1, Bologna, 40127, Italy  \\
ioannis.georgantopoulos@oabo.inaf.it}

%

\maketitle


\begin{abstract}
I present a review of X-ray and mid-IR surveys for Compton-thick Active 
 Galactic Nuclei (AGN). These are the most highly 
 obscured sources having hydrogen column densities $>1.5\times 10^{24}$ $\rm cm^{-2}$. 
 Key surveys  in the local Universe are presented including the 
  high energy SWIFT/BAT and INTEGRAL surveys, mid-IR and also optical surveys.
  Recently, deep X-ray surveys with Chandra and  XMM-Newton 
   have produced a number of candidate  Compton-thick AGN at higher redshift 
    primarily in the Chandra Deep Field South region. 
     In addition, mid-IR surveys with Spitzer have helped to develop 
     novel complementary techniques for the selection of Compton-thick AGN.  
     The mid-IR techniques used to identify Compton-thick AGN  include:
     a) 24 $\rm \mu m$ excess sources relative to their optical emission 
     b) Spitzer spectroscopy for the detection of high optical depth Si 9.7$\rm \mu m$ 
      absorption features c) low X-ray to 6$\rm \mu m$ luminosity ratio. 
\end{abstract}


\section{Introduction}	
The deep X-ray Universe has been probed at unparalleled depth thanks to the
Chandra mission. The deepest Chandra X-ray surveys revealed an AGN sky
density above 10.000$\rm deg^{-2}$ (Alexander et al. 2003, Luo et al. 2008, Xue et al. 2011), probing 
a flux limit of $f_{2-10\,keV}\approx 5 \times 10^{-17}$\, $\rm erg~cm^{-2}s^{-1}$
 (for a review of X-ray surveys see Brandt \& Hasinger 2005). In contrast, the optical surveys for
QSOs reach sky densities of about few hundred per square degree at a magnitude
limit of $B=22$mag (e.g. Wolf et al. 2003). 
 Therefore, hard (2-10\,keV) X-ray
surveys have provided so far the most powerful method for the detection of AGN.

However, even the efficient hard X-ray surveys may be missing a substantial
fraction of the most heavily obscured sources, the Compton-thick AGN, which
have column densities $\rm >10^{24}\,cm^{-2}$. At these high column densities, 
the attenuation of X-rays is mainly due to Compton-scattering rather
than photoelectric absorption. Although there are only a few dozen of 
Compton-thick AGN identified in the local Universe,
 (for a review see  Comastri 2004), there is concrete evidence for the
presence of a much higher number. The peak of the X-ray
background at 20-30\,keV (Frontera et al. 2007, Churazov et al. 2007, Moretti et al. 2009)
can be reproduced by invoking a significant number of Compton-thick
sources at moderate redshifts. However, the exact density of Compton-thick
sources required by X-ray background synthesis models still remains an open
issue (Gilli et al. 2007, Sazonov et al. 2008, Treister et al. 2009, Ballantyne et al. 2011).
  Additional evidence for the
presence of a numerous Compton-thick population comes from the directly
measured space density of black holes in the local Universe
 (Soltan1982). It is found that this space density is a factor of
1.5-2 higher than that predicted from the X-ray luminosity function
 (Marconi et al. 2004, Merloni \& Heinz 2008), although the exact number depends on the
assumed efficiency in the conversion of gravitational energy to radiation.

In this paper, we present a brief review of Compton-thick AGN. We start from 
X-ray spectroscopic results of nearby compton-thick AGN. We then discuss X-ray surveys 
 for Compton-thick sources in both the local universe (z$<$0.1)  
 as well as deep X-ray surveys which probe moderate redshifts where the bulk of the 
  X-ray background is produced. 

\section{X-ray spectroscopy of Compton-thick AGN}
When the absorbing column exceeds $\rm 10^{24}$ 
 $\rm cm^{-2}$, the attenuation in X-ray wavelengths is because 
  of electron scattering rather than photoelectric absorption.
   The main characteristics of a Compton-thick spectrum are:
   a) a flat spectrum with $\Gamma \sim 1 $ at energies below 10 keV.
    This originates in reflection on the back-side of the obscuring screen
     (George \& Fabian 1991, Matt et al. 2004, Murphy \& Yaqoob 2010).
  The flat spectrum is due to the combination Compton down-scattering of high 
   energy photons to lower energies and subsequent photoelectric absorption.   
     b) an $FeK\alpha$ line with large equivalent width, usually of the 
      order of 1keV (e.g. Fukazawa et al. 2011); note however, that 
       there are some exceptions of  Compton-thick AGN with a small $FeK\alpha$ 
        equivalent-width: e.g. Mrk-231  with an equivalent width of
         $\sim 300$eV (Braito et al. 2004). c) An absorption turnover at  
       energies above 10 keV,  with the exact cut-off energy depending on the
        column density. 
       NGC1068 represents an extreme case study of a Compton-thick AGN. 
             This is one of the nearest AGN at a redshift of $z=0.038$. It has a column density 
         of about $10^{25}$  $\rm cm^{-2}$ as derived from BeppoSAX observations 
          (Matt et al. 2004).   At such
           high column densities ($>5\times 10^{24}$ $\rm cm^{-2}$),  
        the transmitted component is totally blocked by the obscuring material.
         The observed X-ray flux, arising from the back-side of the obscuring screen, 
          is diminished by almost two orders of magnitude.
          NGC3079 is  another example of a Compton-thick AGN.  This is mildly Compton-thick 
            having a column density of  $\sim2\times10^{24}$ $\rm cm^{-2}$ (Akylas \& Georgantopoulos 2009). 
           For a detailed discussion of various Compton-thick AGN broad-band SUZAKU spectra 
           selected by SWIFT/BAT, see Comastri et al. (2010).
            An inventory of BeppoSAX observations of nearby AGN, including 
             many Compton-thick AGN can be found in Dadina (2007).    

\begin{figure*}
\rotatebox{0}{\includegraphics[width=6.5cm]{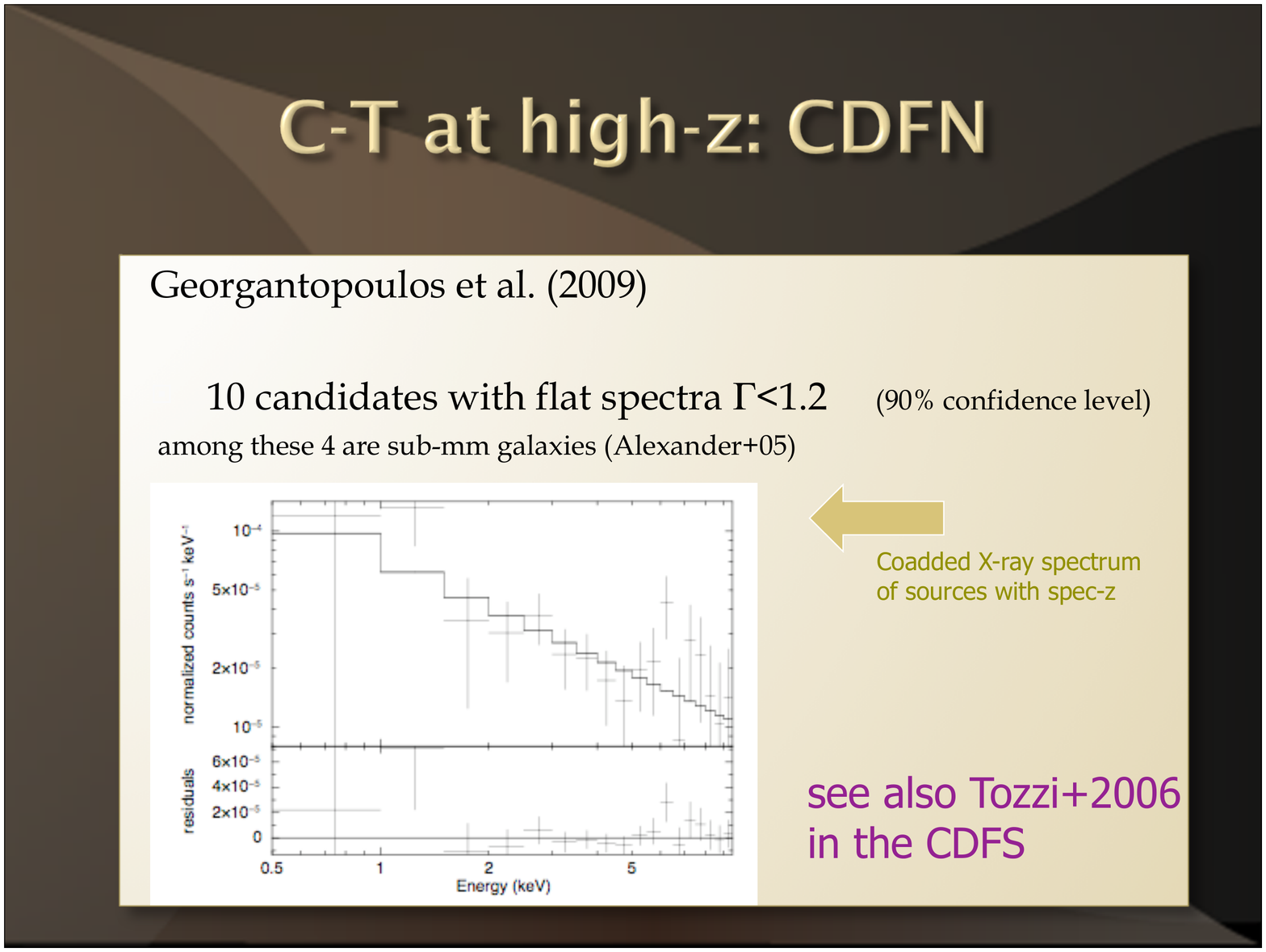}}
\caption{The co-added X-ray spectrum and single power-law fit (as well as residuals) of the six sources in CDF-N with a reflected dominated spectrum 
 and spectroscopic redshifts available (Georgantopoulos et al. 2009). A large equivalent-width ($\sim$1 keV) FeK$\alpha$ line
  at a rest-frame energy of 6.4 keV  can be clearly seen.}
 \label{f1}
 \end{figure*}

\section{Compton-thick AGN in the local Universe}
Comastri (2004) presented a sample from the literature of few tens of Compton-thick AGN 
 in the local Universe. These are classified as Compton-thick AGN 
   on the basis of BeppoSAX observations. Here, we present an update of this review 
     citing the most recent results on the fraction of Compton-thick AGN
      among nearby AGN.
     All the results presented come from direct X-ray spectroscopy,
      either from SWIFT/BAT or from XMM-Newton. 

\subsection{Ultra-hard ($>$10 keV) X-ray selected AGN}
The {\it INTEGRAL} and {\it SWIFT} missions explored the X-ray sky at high energies (15-200 keV),
 providing the most unbiased samples of
Compton-thick AGN in the local Universe. Owing to the limited imaging
capabilities of these missions (coded-mask detectors), the flux limit probed is
 bright  ($\sim 10^{-11}$\, $\rm erg~cm^{-2}s^{-1}$ ), allowing only the detection of AGN at very low
redshifts. These surveys did not detect large numbers of Compton-thick sources
 (Ajello et al. 2008, Tueller et al. 2008, Paltani et al. 2008,Winter et al. 2009, Burlon et al. 2011). The
fraction of Compton-thick AGN in these surveys does not exceed a few percent of
the total AGN population. However, Burlon et al. (2011) point out, that 
 even these ultra-hard surveys are biased against the most heavily
obscured  ($>2\times 10^{24}$\, $\rm cm^{-2}$) Compton-thick sources.

\subsection{Mid-IR selected AGN}
Brightman \& Nandra (2011) present XMM-Newton X-ray spectra for 126 sources from 
 the 12$\rm \mu m$ sample of Rush et al. (1993). This sample contains 893 mid-IR selected local galaxies
  of which about 10\% are AGN. There are  12 Compton-thick AGN in this sample 
    according to the X-ray spectroscopic criteria, resulting in a fraction of Compton-thick AGN of 
    about 20\%. 

\subsection{Optically selected AGN}
 Akylas \& Georgantopoulos (2009) present XMM-Newton spectral analysis of all 38 Seyfert galaxies from the Palomar spectroscopic
sample of galaxies (Ho et al. 1997).  These are found at distances of up to 67 Mpc and cover the absorbed 2-10 keV luminosity
range  $\rm L_{2-10}\sim 10^{38}- 10^{43}$ $\rm erg~s^{-1}$.  
 Three of these sources  are mildly Compton-thick with column densities just above $\rm 10^{24} cm^{-2}$ 
and high equivalent width FeK$\alpha$ lines ($>$ 700 eV). 
 When one considers only the 21 brightest sources with X-ray luminosity above 
  $L_X=10^{41}$ $\rm erg~s^{-1}$, which are contributing significantly to 
   the X-ray background, the fraction of Compton-thick sources becomes 15\%. 
  The Palomar spectroscopic survey finds AGN at very low redshifts. Attempts to 
   detect Compton-thick AGN at higher redshifts, using the X-ray to [OIII] or [NeV] 
    luminosity ratio have been performed by Vignali et al. (2004) and Gilli et al. (2010).  
   We summarize the comparison of the fraction of Compton-thick AGN, in the local Universe, 
    from surveys in different wavelengths in table \ref{tab1}.

\begin{table}[ph]
\tbl{Compton-thick AGN fraction in the local Universe}
{\begin{tabular}{@{}ccc@{}} \toprule
 Survey & CT fraction & Ref \\  \hline 
  IRAS-12$\rm \mu m$  &     0.2    &  Brightman \& Nandra 2011   \\
  SWIFT &     0.04 &   Burlon et al. 2011 \\
  Palomar&     0.15  &  Akylas \& Georgantopoulos 2009\\  \botrule
\end{tabular} \label{tab1}}
\end{table}

\section{Deep X-ray surveys}
\begin{figure*}
\rotatebox{0}{\includegraphics[width=9.0cm]{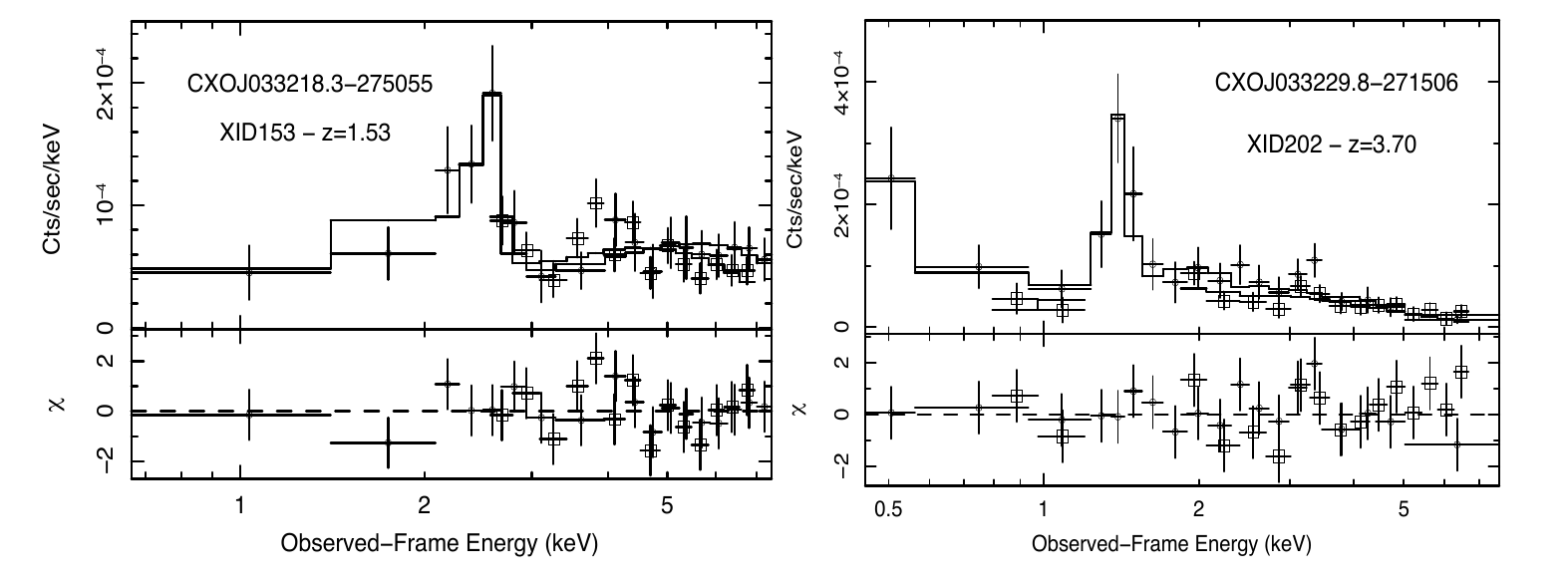}}
\caption{The XMM-Newton spectra of  the two high-redshift Compton-thick sources 
presented in Comastri et al. (2011). }
 \label{f2}
 \end{figure*}

The deep surveys with Chandra and XMM provided the opportunity 
 to search for Compton-thick AGN, mainly through X-ray spectroscopy,  
 at faint fluxes and moderate redshifts ($z\sim0.7-1$) 
where the bulk of the X-ray background is produced. 
The population synthesis models (Gilli et al. 2007) predict an upturn in the number 
 of Compton-thick AGN at fluxes of $\sim 10^{-15}$ $\rm erg~cm^{-2}~s^{-1}$ (2-10 keV) 
  about a factor of 20 brighter than the  flux limit of the Chandra Deep Field South 
   (CDF-S) 4Ms survey (Xue et al. 2011). Tozzi et al. (2006) and Georgantopoulos 
    et al. (2007) first applied X-ray spectroscopy in the 1Ms CDFS observations 
     to identify  reflection-dominated sources finding 
     a couple of tens candidate Compton-thick sources. Georgantopoulos et al. (2009) 
      repeated this exercise in the CDF-N. The above authors find 10 candidate Compton-thick 
        sources down to a flux of $\rm f_{2-10}\sim 10^{-15}$ $\rm erg~cm^{-2}~s^{-1}$. 
         The co-added spectrum of the six sources with available spectroscopic 
          redshifts presents a FeK$\alpha$ line at 6.4 keV rest-frame energy 
           with an equivalent width of $\sim$1keV (Fig. \ref{f1}) showing unambiguously 
            that a large fraction of these are associated with Compton-thick sources.  
       Comastri et al. (2011), using the 3Ms XMM-Newton observations in the CDF-S, re-examined 
        the candidate Compton-thick sources found by Tozzi et al. (2006). 
        They find at least two bona-fide Compton-thick sources, one at a redshift 
         of z=1.53 and another one at z=3.7 (see also Norman et al. 2002), 
          see Fig. \ref{f2}.
         Feruglio et al. (2011) and Gilli et al. (2011) analysing Chandra data in the CDF-S 
         find another two Compton-thick sources at  high redshift.  The one found by Gilli et al. 
          (2007), at z$\sim5$, is the highest redshift AGN discovered so far.
           Recently, Brightman \& Ueda (2012) explore the X-ray spectroscopic 
              of high redshift AGN in the Chandra Deep Field-South (CDFS), making use of  new X-ray spectral models 
              which account for
 Compton scattering and the geometry of the circumnuclear material. They find forty Compton-thick
  candidates.  
               

\section{Mid-IR surveys}
The mid-IR wavelengths have attracted much attention for providing an
alternative way to detect heavily obscured systems. This is because the
absorbed radiation by circumnuclear dust is re-emitted in the IR part of the
spectrum (Soifer et al. 2008), rendering heavily obscured AGN copious mid-IR
emitters. In such systems, the 2-10\,keV X-ray emission can be diminished 
by almost two orders of magnitude (Matt et al. 2004), while at the same
time the isotropic mid-IR emission remains largely unattenuated. 

\subsection{X-ray to mid-IR luminosity ratio}
The presence of a low $\rm  L_X/L_{6\mu m}$ luminosity ratio has
been used as the main instrument for the selection of faint
Compton-thick AGN which remain undetected in X-rays
(e.g. Alexander et al., 2008; Goulding et al., 2011). This is
because the 6$\rm \mu m$ luminosity is considered as a good proxy of the
AGN power, as it should be dominated by very hot dust
heated by the AGN (e.g. Lutz et al., 2004). 
 At these wavelengths, the contribution of the stellar
light and colder dust heated by young stars should be 
 minimal. 
 
\begin{figure*}
\rotatebox{0}{\includegraphics[width=6.3cm]{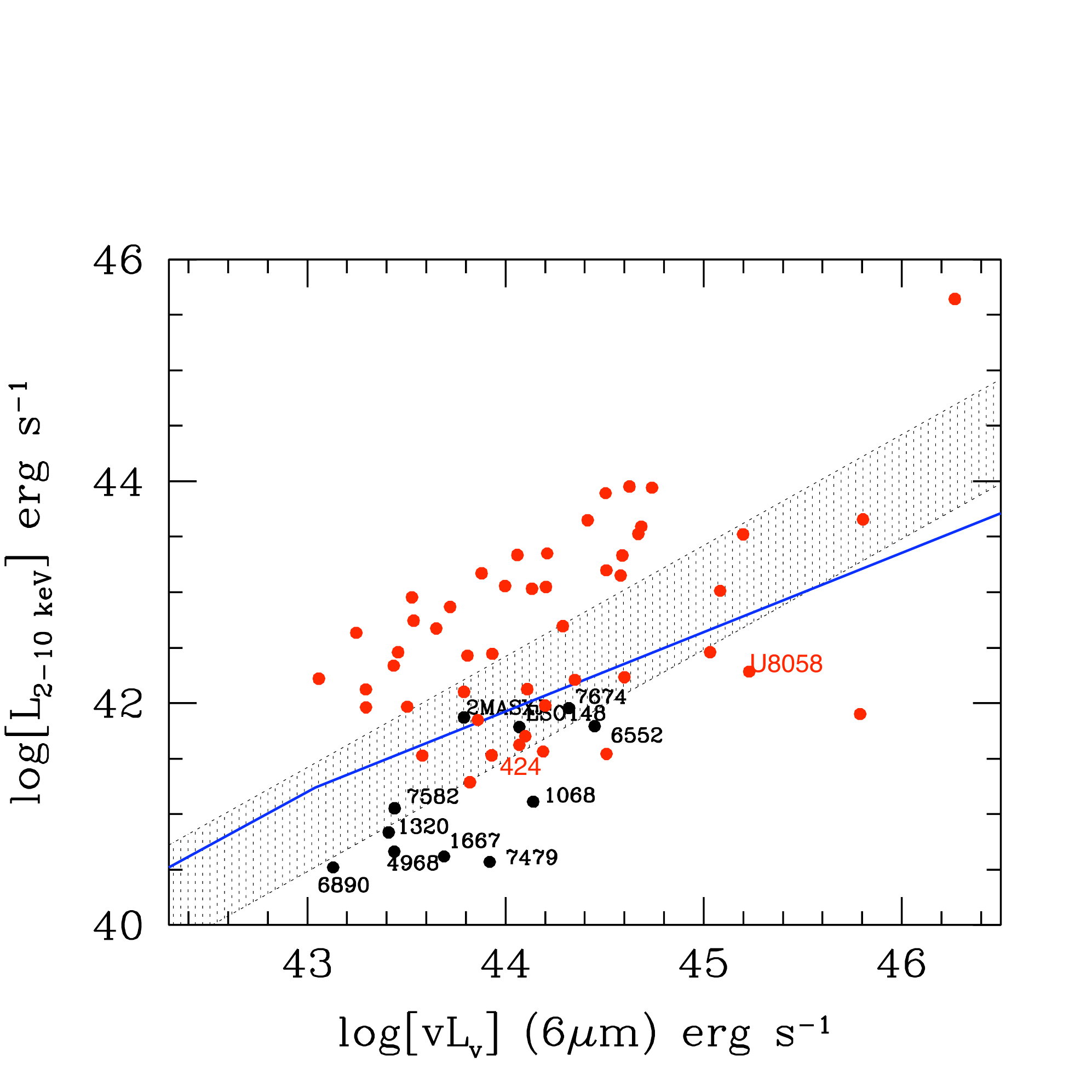}}
 \rotatebox{0}{\includegraphics[width=6.3cm]{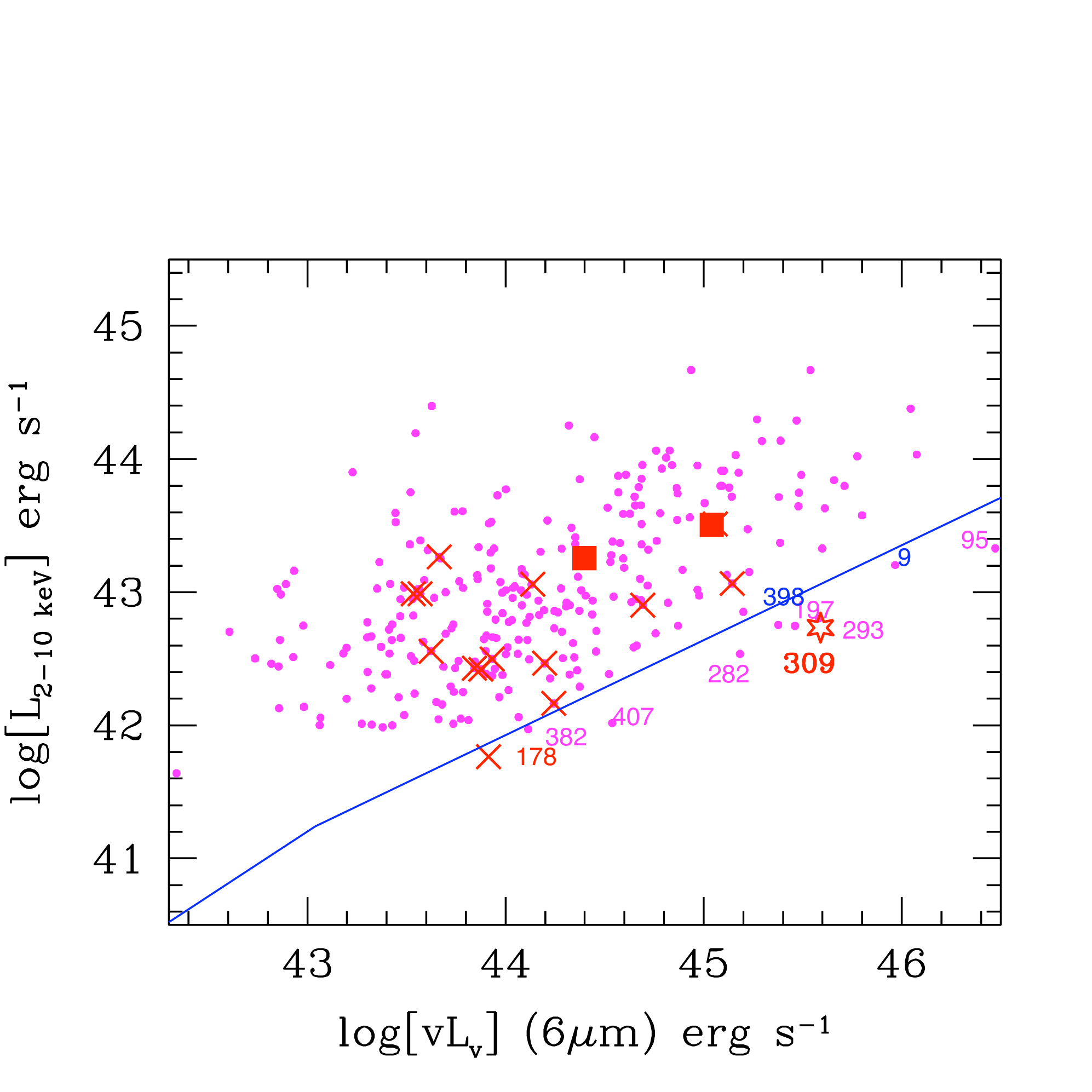}}\hfill
\caption{{\bf Left panel}: The X-ray (2-10 keV) vs 6$\rm \mu m$ luminosity ratio 
 for the IRAS 12$\rm \mu m$ sample of Brightman \& Nandra (2011). The black points 
  denote the AGN which are unambiguously associated with Compton-thick sources 
   on the basis of XMM-Newton spectroscopy. The blue line corresponds to the COSMOS 
    average AGN $\rm L_X/L_{6\mu m}$ luminosity ratio scaled down to  3\%  
     to account for the attenuation of X-rays in Compton-thick AGN. The hatched area 
      corresponds to the local AGN X-ray to mid-IR luminosity relation (Lutz et al. 2004) 
       and its associated 1$\sigma$ error reduced down to 3\%. {\bf Right panel}: The filled circles 
        denote the X-ray
        selected AGN in the CDFS with luminosity $\rm L_X>10^{42}$ $\rm erg~s^{-1}$. 
        in the CDFS. The red squares correspond to the two bona-fide Compton-thick 
         AGN in Comastri et al. (2011). Crosses denote the reflection-dominated Compton-thick 
          AGN in Tozzi et al. (2006) while the open star corresponds to the Compton-thick 
           AGN in Feruglio et al. (2011). For more details see Georgantopoulos et al. (2011d).}
 \label{f3}
 \end{figure*}

Georgantopoulos et al. (2011d) have assessed the efficiency of
the X-ray  ratio method,  estimating the percentage 
of Compton-thick AGN among the low $\rm L_X/L_{6 \mu m}$
sources. They used a sample of local AGN, the IRAS
12$\rm \mu m$ sample of Rush, Malkan \& Spinoglio (1993).
  For this sample excellent quality X-ray spectroscopic observations 
  are available (Brightman \& Nandra, 2011), and hence there is  
   a priori knowledge on which objects are Compton-thick.
  The $\rm L_X/L_{6\mu m}$  diagram for the nearby sources
is shown in Fig. \ref{f3}. The X-ray luminosity is uncorrected for obscuration, 
  while the  6$\rm  \mu m$ luminosity includes both the torus
and the star-formation component. Most (ten
out of eleven) of the sources classified as Compton-thick on
the basis of X-ray spectroscopy lie in the Compton-thick regime. 
 However, many more sources (twelve) that
are not Compton-thick, according to the XMM-Newton X-
ray spectroscopic diagnostics, would be classified as candidate
   Compton-thick AGN on the basis of the low 
$\rm L_X-L_{6 \mu m}$ ratio.
  
  In addition,  the above authors select candidate Compton-
thick AGN among the low $\rm L_X/L_{6 \mu m}$  AGN in the CDF-S.
   In this field both Chandra (4Ms) and XMM (3Ms) data are available,
    rendering it the most well observed area in the X-ray sky. 
  They cannot confirm that a large fraction of the 
   low  $\rm L_X/L_{6 \mu m}$ CDF-S sources are Compton-thick
on the basis of the X-ray spectroscopy. 
   Nevertheless, these sources are highly obscured, having column densities
above $10^{23}$ $\rm cm^{-2}$ . Most importantly, 
 they found that the two bona-fide Compton-thick sources from Comastri et al. (2011)
 do not appear to have a low $\rm L_X/L_{6 \mu m}$ ratio.
   This suggests that the X-ray to mid-IR luminosity 
cannot be used on its own to reliably classify sources as Compton-thick, 
  casting doubt on the efficiency of this method.


\subsection{Mid-IR ($24 \mu m$) excess AGN} 
A number of authors have proposed that sources with  mid-IR 
 excess at 24$\rm \mu m$ are hosting heavily obscured and probably 
  Compton-thick nuclei. Martinez-Sansigre et al. (2005) first argued that 
   a population of bright 24$\rm \mu m$ AGN with no 3.6 $\rm \mu m$ detection 
    is as numerous as unobscured QSOs at high redshift z$>$2.
     Houck et al. (2005)  have detected a population of 
      24 $\rm \mu m$ bright sources that are very faint in the optical 
       R-band having $\rm f_{24\mu m}/f_R> 1000$. These sources have been 
        nicknamed DOGs (Dust Obscured Galaxies). The redshift of these sources 
        is z$\approx$2   with a small scatter $\sigma_z\approx 0.5$  (Dey et al. 2008, Pope et al. 2008).
         Fiore et al. (2008), Georgantopoulos et al. (2008), Fiore et al. (2009), 
         Treister et al. (2009b), Eckart et al. (2010) argue that the stacked X-ray signal 
          of the 24-$\rm \mu m$ infrared excess AGN  are associated with heavily obscured 
           AGN below the flux limit of the deepest Chandra surveys (see also Daddi et al. 2007, 
            Luo et al. 2011). These sources 
            have a stacked signal with a spectrum  of $\Gamma\sim 1$. 
             Georgakakis et al. (2010), examine the X-ray spectra of a number 
              of low redshift ($z\approx 1$) analogs of DOGs, in the AEGIS and CDF-N surveys. 
               They find that only three galaxies may show tentative evidence 
                for Compton-thick obscuration.
           Finally, Georgantopoulos et al. (2011b) present an X-ray spectral analysis of 
           all 24-$\rm \mu m$ excess sources found in the Chandra deep fields
            (22 sources). 
          They identify a number of  at most 12  intrinsically flat $\Gamma<1$  sources which 
           could be associated with Compton-thick sources. On the other hand, 
            a large number of the remaining ten sources have steep X-ray spectra ($\Gamma > 1.4$). 
             They conclude that  the fraction of Compton-thick  sources among DOGs cannot 
              exceed roughly 50\% among the 24$\rm \mu m$ excess sources. 
           
\subsection{AGN with large optical depth Si 9.7 $\rm \mu m$ absorption} 
       Observations with the IRS spectrograph onboard Spitzer have found 
          many sources with very deep Si features at $\rm 9.7\,\mu m$, that have 
          optical depths of $\tau>1$. Since it is believed that a few of these
          systems in the local Universe are associated with Compton-thick active galactic nuclei 
          (e.g. NGC4945, Armus et al. 2007), it is interesting to investigate 
           whether the presence of a strong Si
          absorption feature is a good indicator of a heavily
          obscured AGN.  Georgantopoulos et al. (2011c) have compiled 
           X-ray spectroscopic observations, available
          in the literature, on the optically-thick ($\rm \tau_{9.7\,\mu m}>1$)
          sources from the $\rm 12\,\mu m$ IRAS Seyfert sample. They find that
          the majority of the high-$\tau$ optically confirmed Seyferts (six out of nine) in
          the $\rm 12\,\mu m$ sample are probably Compton-thick. Thus,
            there is direct evidence of a connection between mid-IR
          optically-thick galaxies and Compton-thick AGN, with the success rate
          being close to 70\% in the local Universe. 
          However, this technique cannot provide complete Compton-thick AGN
          samples, i.e., there are many Compton-thick AGN that do not display
          significant Si absorption, with the most notable example being
          NGC\,1068. 
          
          In addition, Georgantopoulos et al. (2011c) have
          constructed  a sample of candidate Compton-thick AGN at higher redshifts.
          This consists of seven high-$\tau$ {\it Spitzer} sources in the 
          Great Observatories Origins Deep Survey (GOODS)
          and five in the {\it Spitzer} First-Look Survey. All these have been
          selected to have no  PAH features (EW$_{6.2\mu m}<$0.3$\rm \mu m$) to maximise the
          probability that they are bona-fide AGN. Six out of the seven GOODS sources
          have been detected in X-rays, while for the five FLS sources only X-ray flux upper limits are available.
           The high X-ray luminosities ($\rm L_X>10^{42}$ $\rm erg~s^{-1}$) of the detected GOODS sources 
           corroborates that these are AGN. For FLS, ancillary optical spectroscopy reveals hidden nuclei in
         two more sources. SED fitting can support the presence of an AGN in the vast majority of  sources.
          Unfortunately, owing to the limited photon statistics, 
          no  useful constraints can be derived from X-ray spectroscopy.
           Therefore the efficiency of the high-$\tau$ method for finding Compton-thick 
            AGN at high redshift remains uncertain. 
            
            \subsection{Compton-thick AGN in sub-mm galaxies}
            The first extragalactic surveys at 850 $\rm \mu m$ with the SCUBA 
             detector on the JCMT telescope revealed a population of
numerous luminous high redshift sub-millimeter galaxies or
SMGs (see Blain et al. 2002, Maiolino 2008). Sub-mm surveys
are very efficient in detecting distant galaxies because of the
negative K-correction at sub-mm wavelengths which counteracts the dimming of light with 
 increasing distance.  Chapman et al. (2005)  have
shown that the SMG population lies at high redshift with
the median being z=2-3,  although the highest redshift SMGs have 
been found out to  $z=4-5$ (Capak et al. 2008, Wardlow et al. 2011).  
 Mid-IR spectroscopy reveals that the SMGs generate large fractions of their energy 
 due to star-forming processes (Pope et al. 2008, Menendez-Delmestre 2009). 
 Deep X-ray surveys  allow the  assessment of the AGN content of distant SMGs. 
  Alexander et al. (2005a, 2005b) have found X-ray
counterparts to a sample of 20 SMGs with radio counterparts from Chapman et al. (2005) in the Chandra Deep
Field North (CDFN). Based on the number of X-ray detections, Alexander et al. (2005b) claim a high fraction
(75$\pm$19 \%) of AGN among the sub-mm galaxies which
have radio counterparts. If one takes into account the radio undetected SMGs in the CDFN, making the 
 conservative assumption that none of the radio undetected SMGs
hosts AGN activity, then the AGN fraction in the SMG
population becomes $>$ 38$^{+12}_{-10}$  \% (Alexander et al. 2005).
Alexander et al. (2005b) argue that the vast majority of the
radio-detected SMGs are highly obscured with column densities 
 exceeding $10^{23}$  $\rm cm^{-2}$. The above results suggest that the 
  intense star-formation goes hand-in-hand with supermassive black-hole growth. 
  Laird et al. (2010) presented an analysis of 35 SMGs in the CDFN
with sub-arcsec positions either from radio or Spitzer counterparts. 
 Their X-ray spectra and low X-ray  luminosities 
    reveal that a dominant AGN contribution is only required in only
seven sources, or 20\% of the SMG sample.  
 Georgantopoulos et al. (2011b) studied the X-ray properties of the 126 sub-mm sources
of the LABOCA survey in the area of the CDF-S and
extended CDF-S. They find 14 sources detected in X-rays.  
The X-ray luminosities and spectra suggest that most
of the sub-mm - X-rays associations (10/14) host an
AGN while in four sources the X-ray emission could possibly
originate from star-forming processes.
 Hence, the fraction of X-ray AGN among the LABOCA SMG
sample in the area of the CDF-S is at most  26$\pm9$\%.
Only six of the X-ray sources show large amounts of
absorption, $\rm N_H>10^{23}$ $\rm cm^{-2}$, 
 while there is no unambiguous evidence that 
 any of these is associated with a Compton-thick source. 
  Finally, they argue that Compton-thick AGN do not even lie  among the 
   SMGs undetected at X-ray wavelengths. X-ray stacking analysis of the undetected SMGs reveals
a signal with a relatively soft spectrum which  is more
suggestive of a SFR galaxy population.

\end{document}